\documentclass[12pt,a4paper]{article}

\usepackage[utf8x]{inputenc}
\usepackage[T1,T2A]{fontenc}
\usepackage[russian,english]{babel}

\usepackage[a4paper,top=2.5cm,bottom=2.5cm,left=2.5cm,right=2.5cm,marginparwidth=1.75cm]{geometry}

\usepackage{amssymb,amsmath}
\usepackage{slashed}
\usepackage{mathtools}
\usepackage{feynmf}

\usepackage{mathrsfs}
\usepackage{graphicx}
\usepackage{simpler-wick}

\usepackage[export]{adjustbox}
\usepackage[colorinlistoftodos]{todonotes}
\usepackage[colorlinks=true, allcolors=blue]{hyperref}
\usepackage{hyperref}
\usepackage{authblk}
\DeclareMathOperator{\Lagr}{\mathscr{L}}
\usepackage{tikz}
\usepackage{subfigure}
\usepackage{indentfirst} 
\usepackage{cite}
\usepackage{gensymb}

\graphicspath{{Pics/}}

\def\pc{{\, \rm pc}}

\title{Indirect effects of dark matter}
\author[]{K.M. Belotsky}
\author[]{E.A. Esipova}
\author[]{A.Kh. Kamaletdinov}
\author[]{E.S. Shlepkina}
\author[1,*]{M.L. Solovyov}
\affil[]{National Research Nuclear University MEPhI (Moscow Engineering Physics Institute), 115409, Kashirskoe shosse 31, Moscow, Russia}
\affil[*]{\small E-mail: max07s@mail.ru}
\date{}

\begin{document}

\maketitle

\noindent\rule{\textwidth}{1pt}

\begin{abstract}
Here we briefly review possible indirect effects of dark matter (DM) of the Universe. It includes effects in cosmic rays (CR): first of all, the positron excess at $\sim$ 500 GeV and possible electron-positron excess at 1-1.5 TeV. We tell that the main and least model-dependent constraint on such possible interpretation of CR effects goes from gamma-ray background. 
Even ordinary $e^+e^-$ mode of DM decay or annihilation produces prompt photons (FSR) so much that it leads to contradiction with data on cosmic gamma-rays.
We present our attempts to possibly avoid gamma-ray constraint. They concern peculiarities of both space distribution of DM and their physics. The latter involves complications of decay/annihilation modes of DM, modifications of Lagrangian of DM-ordinary matter interaction, and inclusion of mode with identical fermions in final state. In this way, no possibilities to suppress were found yet except, possibly, mode with identical fermions. While the case of spatial distribution variation allows
achieving consistency between different data.
Also we consider stable form of dark matter which can interact with baryons. We show which constraint such DM candidate can get from damping effect in plasma during large scale structure formation in comparison with other existing constraints.\\[0.5cm]
\end{abstract}

\noindent\rule{\textwidth}{1pt}

\section{Introduction}
There is no necessity to explain how important the dark matter problem is in physics and cosmology. Many active attempts on both direct and indirect searches for its candidate are undertaken. Though, indirect searches can hardly give decisive solution of the problem, but they, as a rule, are based on the search for signal from DM virtually from all cosmic space where DM resides.

Cosmic ray (CR) puzzles are a popular subject of such the investigations that have been carried out for a long time now. They began with the first WIMP candidate \cite{Zeldovich}, which was used for interpretation of CR data \cite{Konoplich}. Starting from the end of 20th century, extremely huge number of both experimental and theoretical works explore this topic \cite{PhysRevLett.63.840, PhysRevD.42.1001, PhysRevD.43.1774}.
They provide an essential and very rare information on DM.

But unfortunately, such attempts acquire now just only new constraints. It forces theoreticians to invent more and more sophisticated models.
Here we give one of our latest limitations on explanation of lepton anomalies in CR, that in their time induced a hype. Also we provide our attempts of searching ways to circumvent them. They consist in both specific space DM distribution and interaction physics of DM. The latter in its turn includes specifics both in kinematics of different decay modes and in Lagrangian of DM particles. Though explicit positive answer has not yet been found in the last case, but negative result is also useful.

We consider also another type of DM candidate -- particles which are able to interact with baryons. It is conditionally named SIMP (Strongly Interacting Massive Particles), though a strength of force it possesses is not obligatory strong. We obtain, in simple manner, the constraint coming from large scale structure (LSS) which can turn out to be strongest in some range of SIMP mass. It is applied then to some specific models.

Other possible indirect constraints are shortly reviewed.

Structure of the article is as follows. In Sec.~\ref{CR} we consider possible signal from DM in CR,  then in Sec.~\ref{Solutions} consider possible attempts to circumvent the constraints coming from cosmic gamma-background hindering DM interpretation of lepton excess in CR, then in Sec.~\ref{SIMP} consider constraint coming from LSS, in Sec.~\ref{Other} other effects and in Sec.~\ref{Conclusion} conclusion.

\section{Possible signals in cosmic rays}
\label{CR}
\subsection{Positron anomaly of Pamela and AMS-02}

The rise of the positron fraction at energies of around 100 GeV, first measured by PAMELA experiment \cite{Adriani:2008zr} and then confirmed with high accuracy by AMS-02 \cite{PhysRevLett.110.141102,PhysRevLett.113.121101}, turned a new leaf in the indirect searches of DM. The excess of positrons, named "the positron anomaly", could not be described via classical ways, and DM was a tempting candidate on the role of its source. The discovery gave rise to plethora of models with decaying or annihilating DM particles accounting for this phenomenon. The alternative hypotheses include the modified secondary CR production and diffusion, the pulsars as the new source of positrons and particularity of near-Earth space caused by nearby supernova explosion \cite{2009PhRvL.103e1101Y, PhysRevD.52.3265, 2019PhRvD..99j3022R, 2018PhRvD..97f3011K, 2019arXiv190206173L}.

Such models, if confirmed, can become the first non-gravitational evidence of the DM. What is more, the properties of the DM are still unknown, so one can make a wide range of assumptions. However, models of decaying or annihilating Dark Matter particles are still subject to the set of constraints, such as ones set by CMB (Cosmic Microwave Background) or gamma-rays from Dwarf Galaxies \cite{2015ICRC...34...14C}. 

\begin{figure}[h]
    \subfigure[]{
    \includegraphics[width=0.49\textwidth]{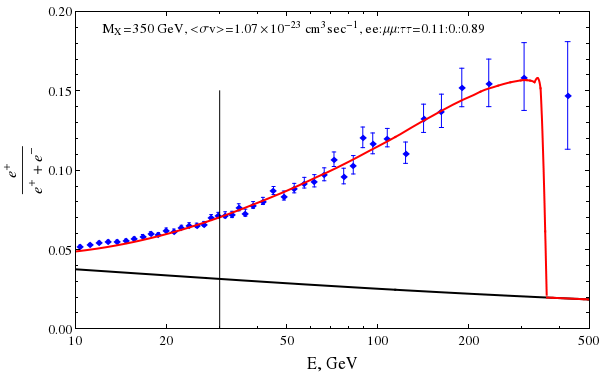}
    \label{halo_e}}
    \subfigure[]{
    \includegraphics[width=0.49\textwidth]{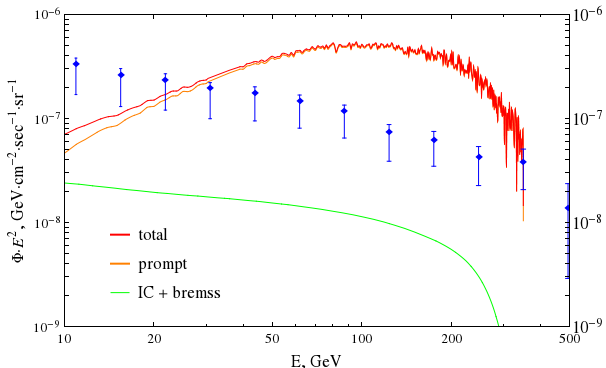}
    \label{halo_g}}
    \caption{The positron fraction  fit \subref{halo_e} and corresponding gamma-ray flux in comparison to IGRB  \subref{halo_g} in the Dark Halo case. 
    }
    \label{halo}
\end{figure}

One of the strictest and at the same time most underrated constraint is set by data on the cosmic gamma-ray background (Isotropic Gamma-Ray Background, IGRB) \cite{Ackermann:2014usa} by Fermi-LAT. It is the background gamma-ray flux averaged over the most of the sky, excluding the Galactic plane. And it has been shown (see Fig.~\ref{halo}), that decaying or annihilating DM distributed in halo produces too many gammas, leading to contradiction with IGRB data \cite{Belotsky:2016tja}.

\subsection{Possible electron-positron excess, registered by DAMPE}

Recent results of DAMPE experiment on the $e^+ + e^-$ spectrum has caused a huge uproar in the scientific community. The most attention is devoted to misaligned datapoint at energy of around 1.5 TeV. Despite the deviation less than $3 \sigma$ \cite{Fowlie:2017fya}, it is often considered as a small line-like signal from DM \cite{Cao:2017rjr,Liu:2017obm,Ding:2017jdr,Li:2017tmd,Chen:2017tva}. However, there are also plenty of works considering the broad excess over the background flux, and it is often viewed as DM signal too (see \cite{2019PDU....2600333B} for references).

\begin{figure}[h]
    \subfigure[]{
    \includegraphics[width=0.49\textwidth]{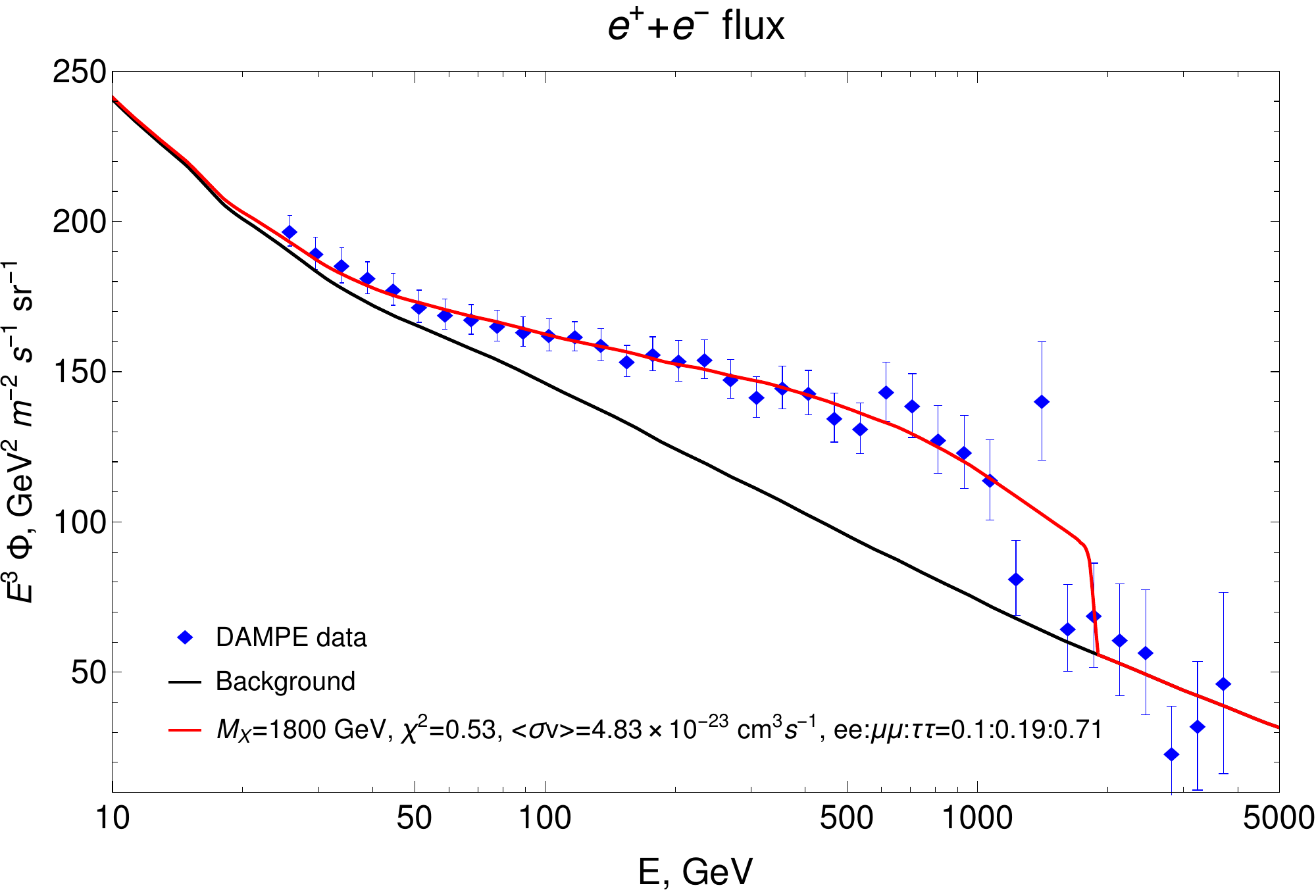}
    \label{dampe_e}}
    \subfigure[]{
    \includegraphics[width=0.49\textwidth]{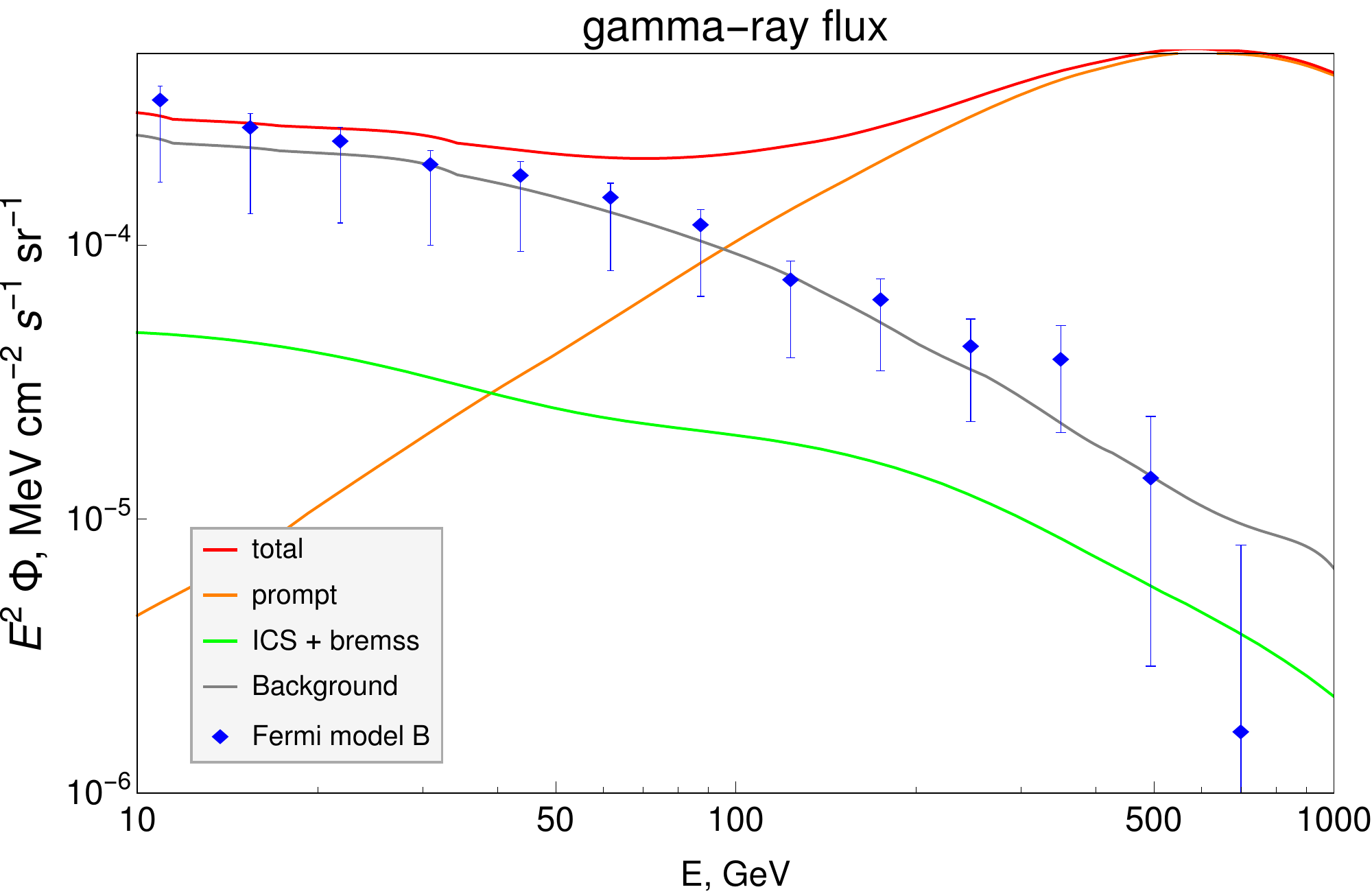}
    \label{dampe_g}}
    \caption{DAMPE $e^+ + e^-$ flux fit \subref{dampe_e} and corresponding gamma-ray flux in comparison to IGRB  \subref{dampe_g}. Note, that $\tau$-channel, the most gamma-producing one, is included. However, its exclusion does not ultimately change the picture.} 
    \label{dampe}
\end{figure}

In our previous work we have studied the possibility of description of this broad excess via the annihilating DM \cite{2019PDU....2600333B}. And it appears the gamma flux is even more of a threat at higher energies. In Fig.~\ref{dampe} we show some of the obtained results.

As can be seen, the main source of the contradiction is high output of direct gamma-rays ("prompt") in the process of decay or annihilation in the form of final state radiation (FSR) and tau-lepton decay products. Therefore, to make description of the anomaly with DM possible, one needs to somehow suppress the gamma-ray production. The possible ways to do so will be discussed further below in our work.

\section{Possible solutions of tensions with gamma-backgrounds and CMB}
\label{Solutions}
\subsection{Resonance DM decay models, some others}

We would like to emphasize that the most model-independent constraint on DM model explaining CR positron (electron) anomaly comes from data on IGRB. The are also constraints based on gamma-rays from dwarf galaxies, galactic center. 

CMB provides independent strong constraint and it, in principle, does not require existence of prompt photons (FSR) in annihilation of DM particles unlike that mentioned above. But they can be avoided but at the same time IGRB constraint still remains in force.
So, how CMB can be evaded:
\begin{itemize}
    \item CMB constraint from Ref. \cite{ade2016planck} on DM model which can explain CR data is applied for annihilating DM only. 
    \item One can try to avoid it, making a tuning with narrow resonance in DM annihilation (e.g.\cite{xiang2017dark,ibe2009breit,bai2018supersymmetric}). Usually it is acheived by tuning velocity of DM particles in so manner that the difference between velocity at CMB epoch and in Galaxy leads to the existence of resonance in Galaxy only. Sometimes CMB (basically in old works) is ignored, and freezing-out and Galactic epochs are considered.
    \item  P-wave annihilation \cite{diamanti2014constraining}, which is proportional to velocity and gives similar effect.
    \item Adjusting two dark species with one decaying (after recombination) into another which annihilates in Galaxy \cite{buch2017late}.
    \item Dark disk.
    \item Or maybe something other.
\end{itemize}

It should be noted one more time that all such attempts along with decaying DM scenario, except maybe dark disk model (which is considered separately below), face difficulty in compatibility with data on gamma-radiation.

Also the long-range self-interacting DM models suggested for solving CR puzzles
cannot avoid CMB data \cite{KB2000,2008PAN....71..147B,2005GrCo...11...27B,2009PhRvD..79a5014A}.

\subsection{Dark disk model}

One of the possible ways to solve the contradiction with cosmic gamma-ray background lies in the spatial distribution of Dark Matter. Due to existence of the magnetic field around the Galactic Disk, positrons born outside of it can not reach the Earth. Meanwhile, gammas are not hindered by it, so these areas still contribute to the gamma-ray flux. Therefore, it is possible to lessen gamma flux without losing in positrons just by "cutting off" area of space outside of the magnetic disk.

In our works \cite{Belotsky:2016tja,Belotsky:2017wgi,1742-6596-675-1-012023,1742-6596-675-1-012026,Alekseev2017An, Belotsky:2018vyt}, we have used this fact to propose the so-called "dark disk model" for description the positron anomaly in AMS-02 data. We consider the Dark Matter to consist of two components. The first component, passive, major one, is supposed to be stable and to form halo. The second, active, minor one, is assumed to form disk and to be able to decay or annihilate with production of $e^+ e^-$ in final state.

In our works we have considered several annihilation models in the framework of the dark disk:
\begin{itemize}
    \item Simple leptophilic model, where DM particles $X$ can annihilate into three leptonic channels $e^+ e^-$, $\mu^+ \mu^-$, $\tau^+ \tau^-$,  with different branching ratios;
    
    \item Cascade leptophilic model, where DM particles can annihilate via the cascade into two lepton-antilepton pairs $X\Bar{X}\rightarrow a\Bar{a} \rightarrow 2 (l^+ l^-)$;
    
    \item Modifications of the two above with extra quark-antiquark channel included.
\end{itemize}

\begin{figure}[h]
    \centering
    \includegraphics[width=0.5\textwidth]{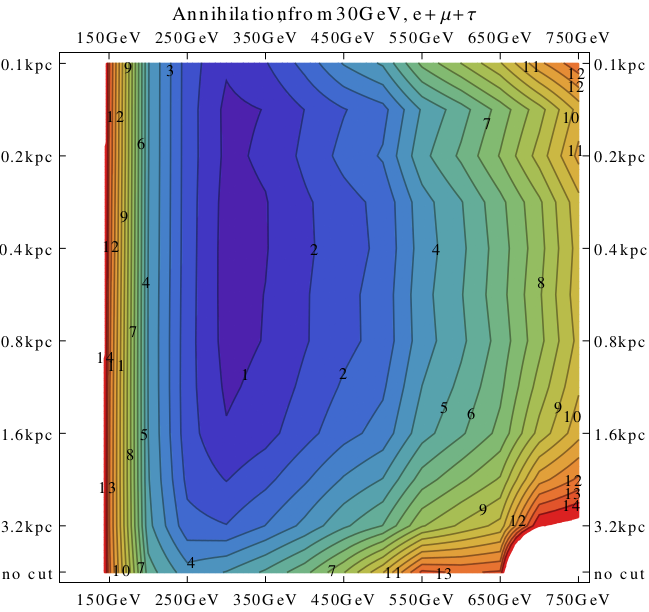}
    \caption{Contour plot for $\chi^2/N_{dof}$ in dependence of disk half-width and mass of DM particle from \cite{Belotsky:2016tja}.}
    \label{disk_width}
\end{figure}

\begin{figure}[h]
    \subfigure[]{
    \includegraphics[width=0.49\textwidth]{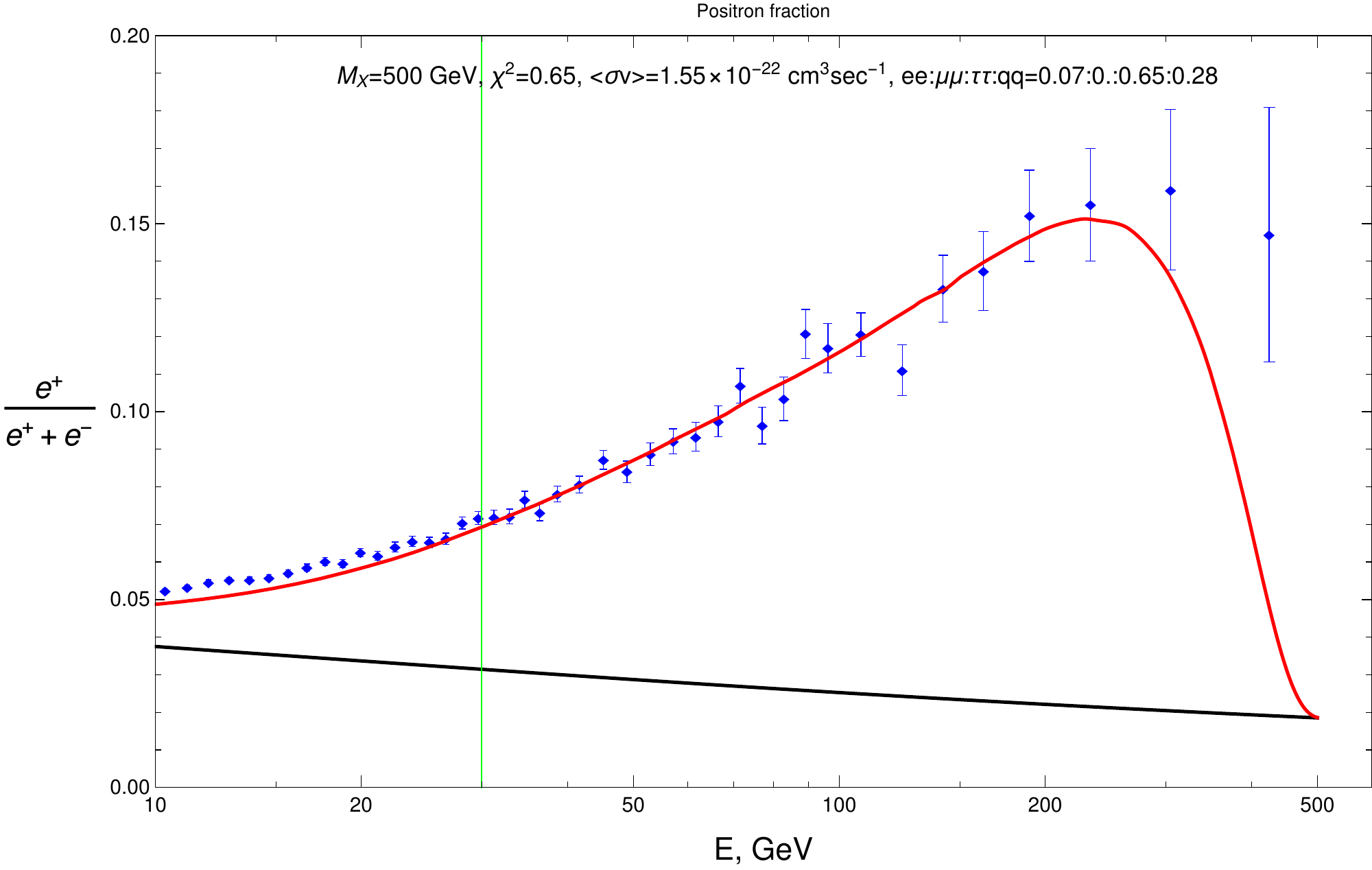}
    \label{disk_e}}
    \subfigure[]{
    \includegraphics[width=0.49\textwidth]{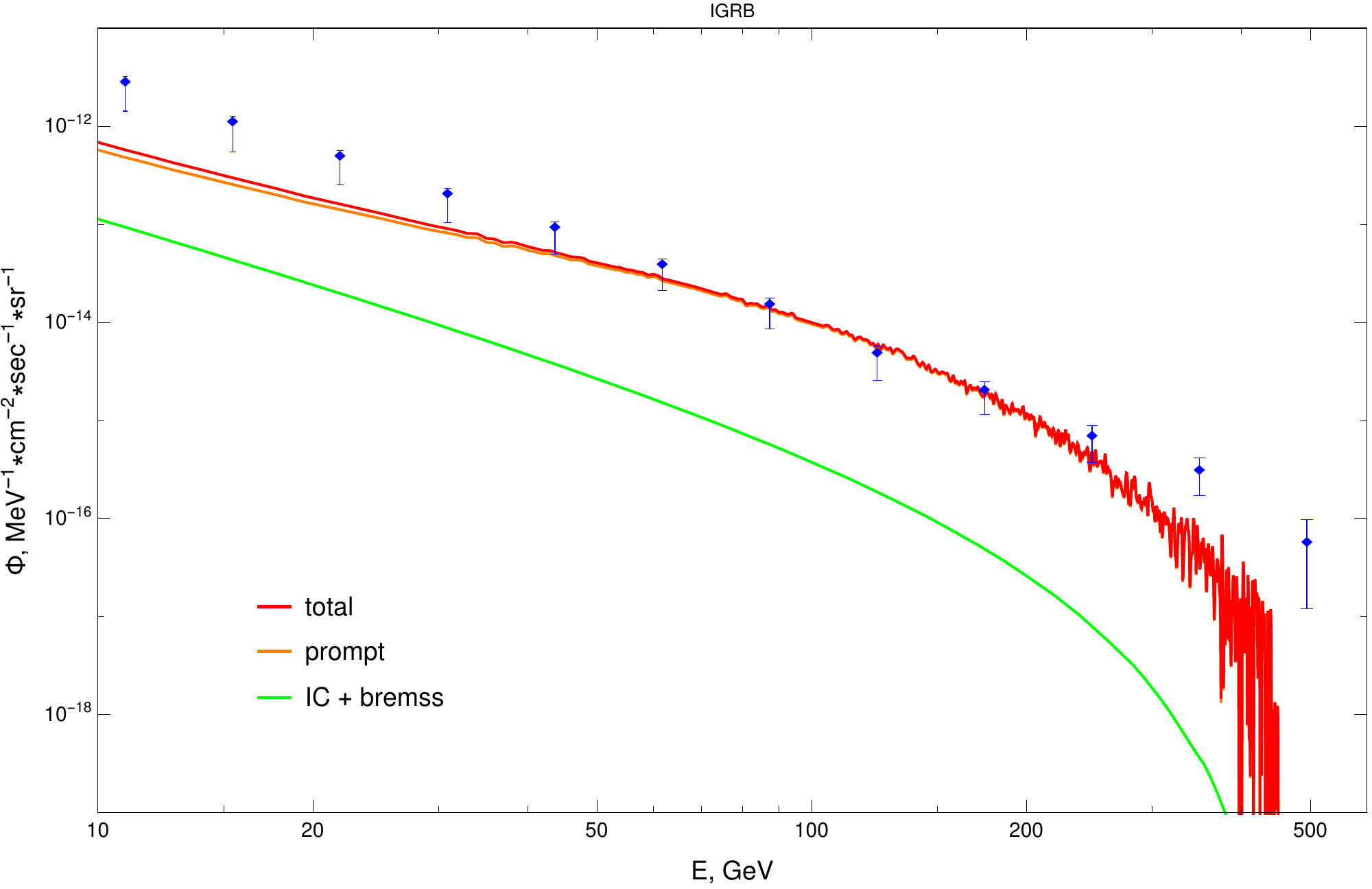}
    \label{disk_g}}
    \caption{Positron fraction \subref{disk_e} and corresponding gamma-ray flux in comparison to IGRB \subref{disk_g} for dark disk with cascade annihilation and additional $2(t\Bar{t})$ channel. Graphics of gamma from Galactic Center and antiproton fraction from original work \cite{Belotsky:2018vyt} are not included.}
    \label{disk}
\end{figure}

Implementation of the dark disk model greatly reduces the contradiction with IGRB data (see Fig.~\ref{disk_width}). Simultaneous fit of positron fraction and gamma flux can also improve the overall goodness of fit at the cost of worsening description of the AMS-02 data. And by usage the complex annihilation models with cascades and quark channels it is possible to achieve value of $\chi^2/N_{dof}\sim 1$ \cite{Belotsky:2018vyt}. In Fig.~\ref{disk} we show some graphics corresponding to this case.

We should also note, that dark disk model naturally avoids the CMB constraints due to active DM component making up a small fraction of all DM. 

In our future works we are going to apply the dark disk model to the DAMPE data. To fit it, one needs much heavier DM particles, than in AMS-02 positron anomaly case, leading to greater high-energy gamma flux and avoiding the contradiction presumably becomes even harder. Therefore the task can give the real test to the model.

\subsection{DM clumps}

\begin{figure}[h]
    \centering
    \includegraphics[width=0.5\textwidth]{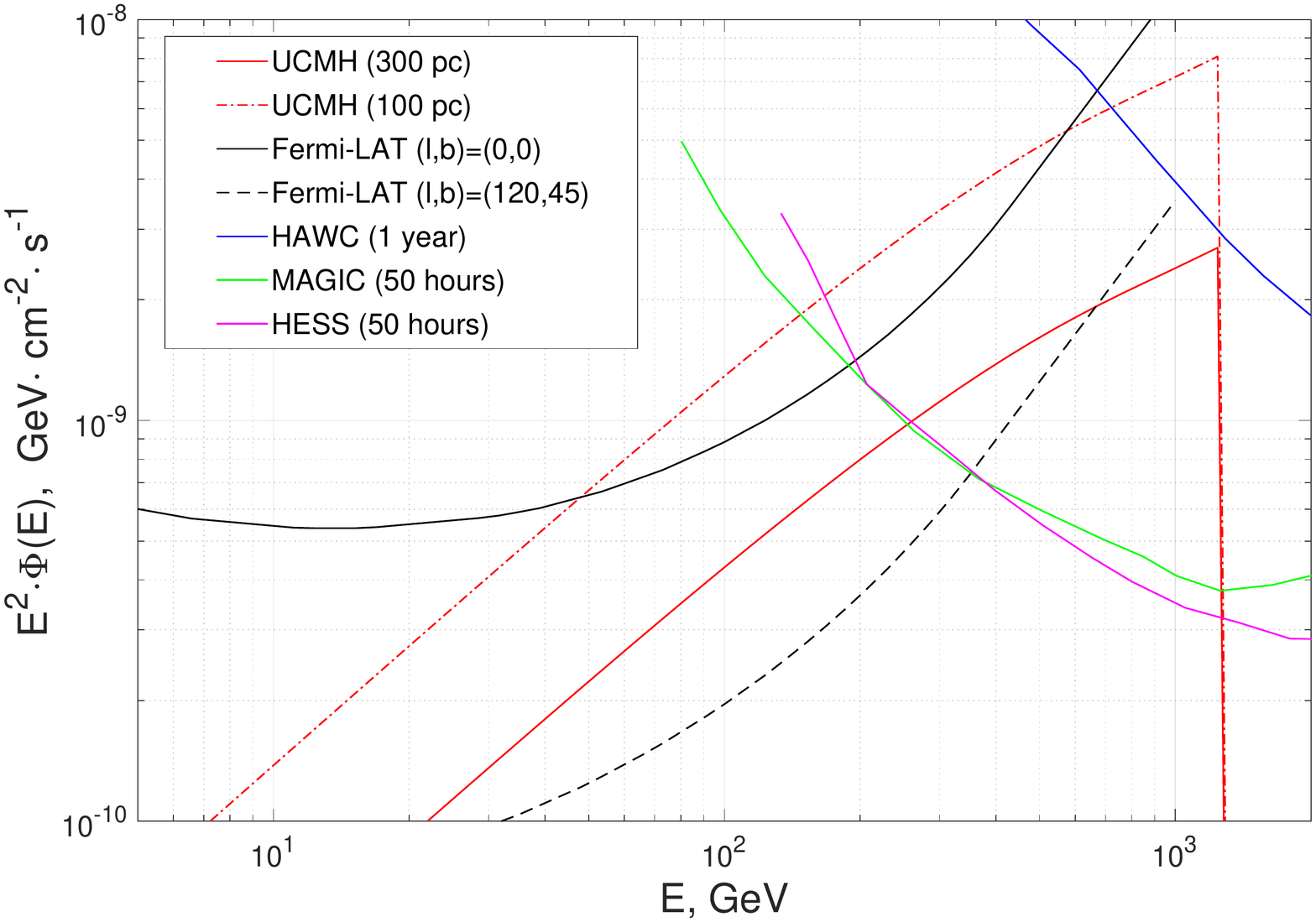}
    \caption{The gamma-ray fluxes from a nearby clump (red solid curve for the distance of $300 \pc$ and red dashed curve for $100 \pc$ respectively) compared to the point-source differential flux sensitivities of different gamma-ray telescopes, including Fermi-LAT (black solid and dashed curves) \cite{FermiLAT:performance}, HAWC (1 year, blue curve), MAGIC II (50 hours, green curve) and HESS (50 hours, magenta curve)~\cite{PointSourceSensitivities}. The black curves show the Fermi-LAT 10-year broadband sensitivities: the solid line corresponds to the minimal sensitivity in the direction of the GC and the dashed line corresponds to the maximal sensitivity in the direction of the Galactic periphery ($l = 120\degree, b = 45\degree$).}
    \label{UCMH_gamma_constr}
\end{figure}

Assumption on existence of a local DM clump is one of the simplest possible explanation of bump in $e^+$ ($e^-$) energy spectrum avoiding constraint from IGRB. It requires tuning (at the cost of low probability) of a distance to the nearest and next to the nearest clumps. 
However, existing gamma ray telescopes (like MAGIC and  HESS) can observe such local clump.
For instance, sensitivity of the major modern gamma-telescopes is already enough to check DM clump model explaining 1.5 TeV datapoint of DAMPE. Respective Fig.~\ref{UCMH_gamma_constr} from \cite{2019PDU....2600333B} is quoted.

\subsection{On FSR suppression due to DM interaction Lagrangian}

\begin{figure}[h]
    \centering
    \includegraphics[width=0.5\textwidth]{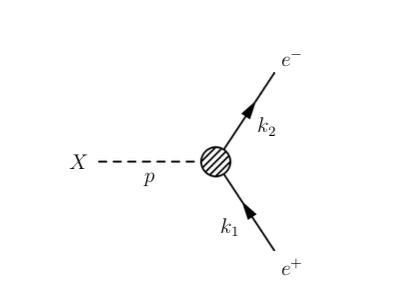}
    \caption{Feynman diagram for two-particle decay of DM particle.}
    \label{feyn1}
\end{figure}

\begin{figure}[h]
    \centering
    \includegraphics[width=0.7\textwidth]{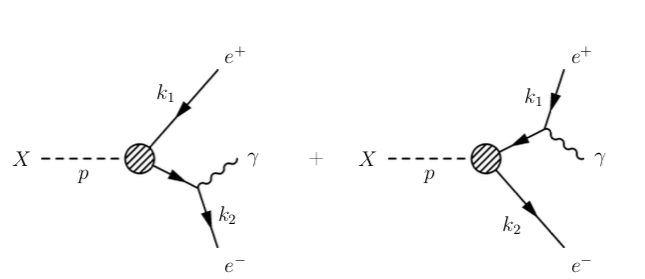}
    \caption{Feynman diagram for three-particle decay of DM particle.}
    \label{feyn2}
\end{figure}

The theoretically predicted photon yield accompanying the formation of leptons can be reduced owing to Dark Matter interactions physics. Here we focus on decays $X\rightarrow~e^+~e^-$ and $e^+ e^+$ taking into account FSR. Typical diagrams (for $e^+ e^-$) of the corresponding processes 
are shown in the Fig.(\ref{feyn1}, \ref{feyn2}).

DM particle $X$ are supposed to be of scalar, pseudo-scalar, vector, and axial-vector type.
The corresponding interaction Lagrangians are: 
\begin{equation}
    \Lagr_\text{s} = X\overline{\psi}\psi, \quad \Lagr_\text{ps} = X\overline{\psi}\gamma^5\psi, \quad \Lagr_\text{v} = \overline{\psi}\gamma^\mu \psi X_\mu, \quad \Lagr_\text{av} = \overline{\psi}\gamma^\mu \gamma^5 \psi X_\mu
    \label{lagrs}
\end{equation}

\subsubsection{The case of $e^+ e^-$-mode}

The idea is to combine the scalar and pseudo-scalar couplings (with some coefficients $a$ and $b$) in the hope of destructive interference between them.
The case of vector and axial-vector couplings is analogous. 
So we have
\begin{equation}
\Lagr_{scalar} = X\overline{\psi}(a+b\gamma^5)\psi
\label{scalar}
\end{equation}
\begin{equation}
\Lagr_{vector} = \overline{\psi}\gamma^\mu (a+b\gamma^5)X_\mu \psi
\label{vector}
\end{equation}
to understand which coupling constants ($a$ and $b$) must be chosen in order to suppress a photon.

The suppression of the photon yield implies minimization of the following ratio of two decay modes widths (fig.~\ref{feyn1}, \ref{feyn2})

\begin{equation}
    R = {\Gamma(X \rightarrow e^+ e^- \gamma) \over \Gamma(X \rightarrow e^+ e^-)} \rightarrow \rm min. \newline
    \label{R}
\end{equation}

To obtain it, squared matrix elements ($|M|^2$) were calculated for the two-body and three-body decay. 
They respectively are:

\begin{equation}
|M|^2 = 4(a^2+b^2)(k_1 k_2) \, ; \quad |M|^2 = (a^2 + b^2)\lbrace \ldots \rbrace \, ,
\end{equation}
where $\lbrace \ldots \rbrace$ does not depend on $a$ and $b$.
So, the ratio $R$ will do the same as factor $(a^2+b^2)$ is canceled in \eqref{R}, and therefore 
there is no suppression of radiation coming from the choice of the interaction vertex in this case.
Note, the calculation was performed both manually and using the ME-generator CalcHEP\cite{belyaev2013calchep} providing the cross-check.
\newline

Unfortunately, the situation was found to be the same for vector DM particles. It can be seen from their squared matrix elements: 

\begin{equation}
    |M|^2 = 4(a^2 + b^2)\lbrace \ldots \rbrace \, ; \quad |M|^2 = 14 (a^2 + b^2)\lbrace \ldots \rbrace \, .
\end{equation}

So, the parametrizations of interaction vertexes (\ref{scalar}, \ref{vector})  gives no FSR suppression.
\newline

\subsubsection{The case of $e^+e^+$ mode}

\begin{figure}[h]
    \centering
    \includegraphics[width=0.8\textwidth]{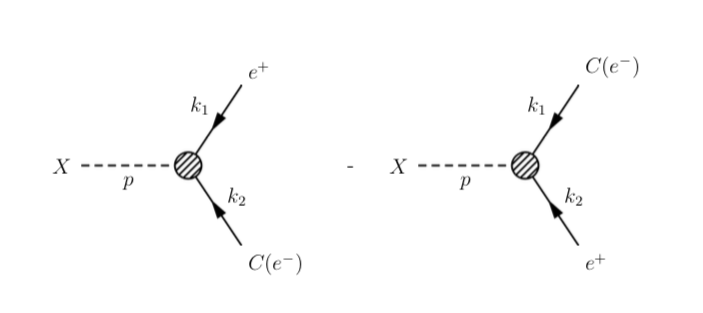}
    \caption{Feynman diagram for two-particle decay of doubly charged DM particle.}
\label{feyn3}
\end{figure}

\begin{figure}[h]
    \centering
    \includegraphics[width=0.8\textwidth]{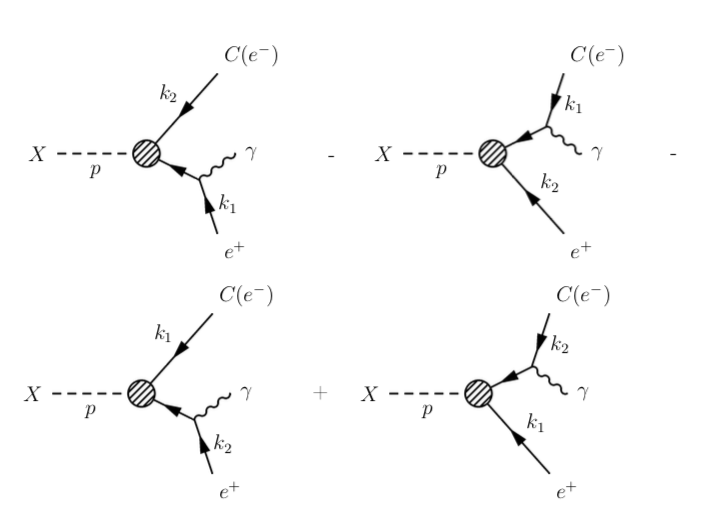}
    \caption{Feynman diagram for three-particle decay of double charged DM particle.}
\label{feyn4}
\end{figure}

Next step which we considered is to take identical fermions in final state, to study decay modes like this: $X \rightarrow e^+ e^+\, {\rm  or   }\; e^- e^-$ (diagrams of the first processes are shown in Fig.\ref{feyn3}). Diagrams of the corresponding process with FSR are given in Fig. \ref{feyn4}.
\newline

One has to do specially several remarks concerning this model case. 
First one is a strangeness of DM particle having a double charge. Nonetheless, such models were suggested in \cite{Khlopov2006,khlopov2006dark,khlopov2008strong,doi:10.1142/S0217751X14430027}, and considered in \cite{Belotsky:2014haa,belotsky2014dark} for possible explanation of positron anomaly. Also one can suppose them to be uncharged considering their cascade decay through some intermediate particles $Y^{++}, Y^{--}$ decaying into lepton pairs.
Next argument in favour of $X^{++}\rightarrow e^+e^+$ mode from viewpoint of FSR suppression is that we simply have one emitted photon per two positrons instead of one positron and one electron.
But there can exist one more argument concerning an identity of final fermions.
This may change influence of Lagrangian parametrization, as well as due to identity of particles themselves. 
The latter is expected implicitly to follow from two arguments. First one comes from the fact that a classical (dipole) radiation of two same signed particles is zero. Second one is partial correspondence of quantum approach with classical one due to so called single photon theorem (or "Radiation Zeros") \cite{brown1995understanding}.
\newline

The Lagrangian of such a model is given below

\begin{equation}
    \Lagr_C = X\overline{\psi^C} (a+b\gamma^5)\psi + X^*\overline{\psi} (a+b\gamma^5)\psi^C ,
\end{equation}
where upper index ~$C$ means charge conjugation.
This theory implies existence particle
and antiparticle ($X^{++}, X^{--}$) corresponding to $X$ and $X^*$ in the Lagrangian.

We assume a case in which the initial state is fixed. That is, there is only particle $X$ in the initial state. Then we can consider only one term of the interaction Lagrangian:
\begin{equation}
    \Lagr_C = X\overline{\psi^C} (a+b\gamma^5)\psi .
\end{equation}

Matrix element are calculated in \hyperref[Appendix]{Appendix}.
The squared matrix elements of the two-body and tree-body decay processes respectively in case of scalar $X$ particle are obtained with the following structure:

\begin{equation}
|M|^2 = 4m^2_{\rm x} (a^2+b^2) \, ; \quad |M|^2 = 16(a^2 + b^2)\lbrace \ldots \rbrace \, .
\end{equation}

And in case of vector $X$ particle, the corresponding squared matrix elements are:

\begin{equation}
|M|^2 = 8m^2_{\rm x} b^2 \, ; \quad |M|^2 = {16b^2\over m^2_{\rm x}} \lbrace \ldots \rbrace \, .
\end{equation}

These results were crosschecked with the help of ME-generator CalcHEP, which allows calculating $|M|^2$ in symbolic form.

So, as in previously considered model cases we do not obtain FSR suppression in both scalar and vector DM particles.
Therefore, it becomes necessary to complicate the models.
\newline

\begin{figure}[h]
    \centering
    \includegraphics[width=0.5\textwidth]{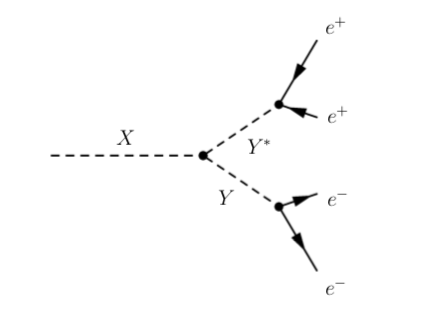}
    \caption{Feynman diagram for the decay of DM particle through a heavy intermediate particle $Y$.}
\label{feyn10}
\end{figure}

One of the models under consideration was the model of dark matter, consisting of scalar or vector uncharged self-conjugated particles $X$, decaying through a massive mediator particle $Y$, which in turn decays into two charged leptons. The Feynman diagram for this process is shown in the Fig.~\ref{feyn10}.

In this case, several variations of such a model can be considered. We simulated and analyzed each such variations using a CalcHEP:

\begin{itemize}
    \item $X$ is a scalar real field, $Y$ is a complex scalar doubly charged field:
 \begin{equation}
\begin{gathered}
\Lagr = XY^*Y + Y\bar{\Psi^C}(a+b\gamma^5)\Psi + Y^*\bar{\Psi}(a-b\gamma^5)\Psi^C - \\ - \bar{\Psi}\gamma^{\mu}A_{\mu}\Psi +
A_{\mu}Y\partial^{\mu}Y^* + A_{\mu}Y^*\partial^{\mu}Y
\end{gathered}
\end{equation}
\item $X$ is a real vector field, $Y$ is a complex scalar doubly charged field:
\begin{equation}
\begin{gathered}
\Lagr = Y\partial^{\mu}X_{\mu}Y^* + X_{\mu}\partial^{\mu}YY^* + X_{\mu}\partial^{\mu}Y^*Y + X_{\mu}A^{\mu}Y^*Y + Y\bar{\Psi^C}(a+b\gamma^5)\Psi + \\ + Y^*\bar{\Psi}(a-b\gamma^5)\Psi^C - \bar{\Psi}\gamma^{\mu}A_{\mu}\Psi + A_{\mu}Y\partial^{\mu}Y^* + A_{\mu}Y^*\partial^{\mu}Y
\end{gathered}
\end{equation}
\end{itemize}

The ratio of the decay widths $\Gamma(X \rightarrow e^+, e^+, e^-, e^-, \gamma) / \Gamma(X \rightarrow e^+, e^+, e^-, e^-)$ in these theories was found to be independent of the parameterization of the interaction Lagrangian.

Thus, none of these models gave a positive result due to Lagrangian parameterization in cases of both non-identical final fermions and the identical ones.

Thus, considered above physical models do not have FSR suppression and therefore are closed ways of the DM interpretations of CR anomaly. 
\newline

Pure effect of identity of final fermions (mode like $X^{++}\rightarrow e^+e^+$), mentioned but not considered here, can be more promising.
The value of ratio
\begin{equation}
    \frac{Br(X\rightarrow e^+e^+\gamma)}{Br(X\rightarrow e^+e^-\gamma)}<1,
\end{equation}
where the $Br()$ is the branching ratio of respective modes,  would be positive result.

\section{CMB and LSS constraints on SIMPs}
\label{SIMP}

\subsection{Generic constraint}
There are many experiments where direct search for strongly interacting dark matter (SIMP) was undertaken. They put constraints on the cross sections of SIMP interaction with baryonic matter. Besides direct constraints there can be indirect ones coming from cosmology.
\newline

Large scale structure of the Universe strongly depends on the moment when DM decouples from ambient relativistic plasma. According to \cite{bertschinger2006effects} and \cite{loeb2005small}, due to Silk-like damping effect\footnote{We were notified by aforementioned authors, that it is not quite Silk effect, nonetheless similar. Silk effect relates to the minimal CMB anisotropy scale, which as a rule is less than horizon.}, the minimal size of inhomogeneities which the Dark matter forms will be determined by the horizon size at the decoupling moment. So we can put a lower limit on the decoupling temperature.
The lower temperature is, the bigger inhomogenities have time to be washed out.
We require that the scales corresponding to dwarf galaxies ($\sim 10^8 M_{\odot}$) should survive. 

Mass of DM inside the horizon depending on the current temperature roughly is $0.1M_{\odot}\left(\frac{T}{1{\rm MeV}}\right)^3$.
So, decoupling temperature 
\begin{equation}
  T_{\rm dec} < 1 \div 5 {\rm  keV} \equiv T_{\rm HDM}
  \label{Tlimit}
\end{equation}
is assumed to be forbidden. Here we denoted the value of upper limit itself as $T_{\rm HDM}$.
This condition is our simplified criterion which we apply here to DM parameters in a broad range.We do not consider the validity of this criterion as it seems to be not very clear.. Somewhere approaches based on free-streaming or diffusion damping effect of primordial inhomogeneities are used (see, e.g.\cite{ooba2019cosmological}) which provide mostly weaker condition. Our goal is to demonstrate that using given simple criterion Eq.\ref{Tlimit} allows obtaining stronger constraint in some parameter regions as compared to existing ones.  

The moment of DM decoupling from baryonic matter can be defined by equation of SIMPs temperature evolution \cite{belotsky2016temperature}
\begin{equation}
    {dT_{\rm s} \over dT} = - {2 \over 3T}\left({1 \over H} \langle \overline{\Delta E} \sigma v \rangle_{\rm sp} n_{\rm p} - 3 T_{\rm s} \right),
    \label{temperature_evolution_eq}
\end{equation}
where
\begin{equation}
\left< \overline{ \Delta E} \sigma v \right>_{\rm{sp}} = 
\frac{8\sqrt{2}}{\sqrt{\pi}} \frac{m_{\rm s} m_{\rm p}}{(m_{\rm s}+m_{\rm p})^2}\sigma_\text{sp} 
\sqrt{\frac{T_{\rm s}}{m_{\rm s}} + \frac{T_{\rm p}}{m_{\rm p}} } \cdot (T_{\rm p} - T_{\rm s}).
\label{simps_equation}
\end{equation}
Here $T_{\rm s}$ and $T=T_{\rm p}$ are the temperatures of the SIMP component of DM and plasma (in Eq.\eqref{temperature_evolution_eq} of photons, in Eq.\eqref{simps_equation} of protons) respectively, $H$ is the Hubble parameter, $n_{\rm p}$ is the concentration of protons (nucleons), $\overline {\Delta E}$ is the kinematically averaged transferred energy between SIMP and baryon, $\sigma$ and $v$ are their cross section and relative velocity respectively, brackets $\left <~ \right> $ mean averaging over thermal distribution.
\newline

\begin{figure}[h]
    \centering
    \includegraphics[width=0.4\textwidth]{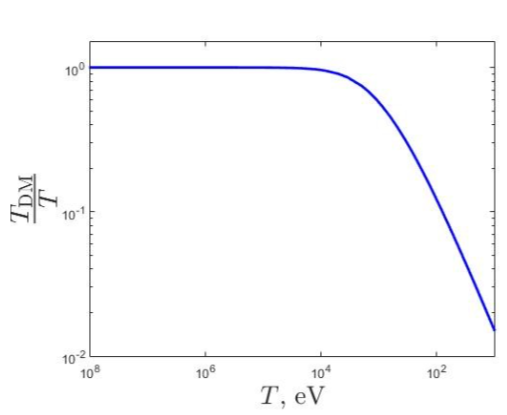}
    \caption{Example of SIMP temperature evolution, obtained from Eq.~\ref{temperature_evolution_eq}}
    \label{bend_figure}
\end{figure}

Decoupling moment was defined as intersection point of two lines tangent to high and low-temperature parts of the SIMP temperature evolution curve shown in Fig.~\ref{bend_figure}. Otherwise, it can be found from equating two terms in Eq.~\ref{temperature_evolution_eq} so we can get roughly (omitting all insignificant coefficients and taking $m_s\gg m_p$)
\begin{equation}
    T_{\rm dec}\sim \left
    (\frac{m_{\rm s}}{\eta_B m_{\rm Pl}m_{\rm p}^{1/2}\sigma_{\rm sp}}\right)^{2/3}.
    \label{Tdec_limit}
\end{equation}

\begin{figure}[h]
    \centering
    \includegraphics[width=0.7\textwidth]{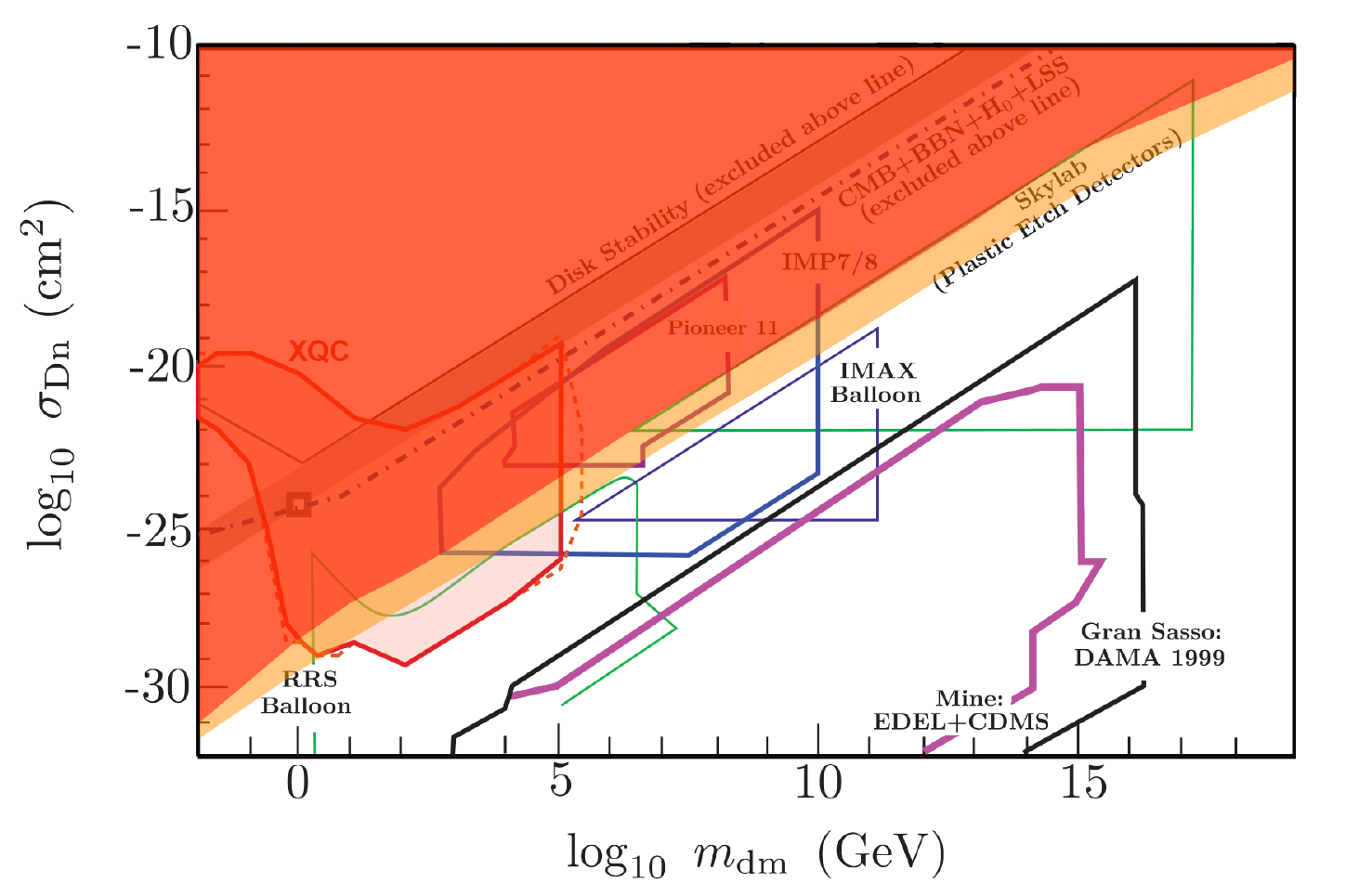}
    \caption{The obtained constraints on the DM-baryon scattering cross section in comparison with limits, based on the DAMIC, XQC and other experiments. Red and yellow regions are our constraints for 5 keV and 1 keV correspondingly.}
    \label{temperature_evolution}
\end{figure}

Condition Eq.\ref{Tlimit} gives us upper limit on allowed $\sigma$.
The result of numerical calculation is shown in Fig.\ref{temperature_evolution} by filled region. Also this can be done symbolically solving roughly Eqs.\ref{Tlimit}-\ref{simps_equation}, what gives
\begin{equation}
    \sigma_{\rm sp}<
    \frac{m_{\rm s}}{\eta_B m_{\rm Pl}m_{\rm p}^{1/2}T_{\rm HDM}^{3/2}}\sim 10^{-29}\text{cm}^2\frac{m_{\rm s}}{\text{GeV}}\left(\frac{5\text{ keV}}{T_{\rm HDM}}\right)^{3/2},
    \label{sigma_limit}
\end{equation}
where $\eta_{B}$ is the baryon-to-photon ratio.

One can see from Fig.\ref{temperature_evolution}, that the used simple criterion can provide the strongest limit on $\sigma$ at some values of $m_s$.

\subsection{Application to particular models}

The obtained constraint on the interaction cross section can be expressed in terms of parameters of specific models. It is not necessary that model should relate to SIMP, but at some values of its parameters it can fall under aforementioned constraint.

First, we take the model of Anthony DiFranzo and Keiko I. Nagao \cite{difranzo2013simplified}.
It contains a fermionic dark matter particle denoted by $\chi$, which is a SM singlet and can be either a Dirac or a Majorana fermion. In addition, this theory contains scalar mediator particles ($\tilde{u}, \tilde{d}, \tilde{q}$) which interact with the dark matter and SM quarks. In this model there are three choices for gauge representations of the mediators under the SM $(SU(3),SU(2))_{Y}$:
\begin{equation}
    \tilde{u}_R = (3,1)_{2/3} , \quad \tilde{d}_R = (3,1)_{-1/3} , \quad \tilde{q}_L = (3,2)_{-1/6}.
\end{equation}

For example, in the model with $\tilde{u}_R$ and Dirac DM the Lagrangian is:
\begin{equation}
    \Lagr = i \bar{\chi} \slashed{\partial} \chi - M_{\chi} \bar{\chi} \chi + (D_{\mu} \tilde{u})^* (D^{\mu} \tilde{u}) - M_{\tilde{u}}^2 \tilde{u}^* \tilde{u} + (g_{DM} \tilde{u}^* \bar{\chi} P_R u + h.c.),
\end{equation}
where $D_{\mu} = \partial_{\mu} - i g_{\rm s} G^a_{\mu} T^a - i {2 \over 3} e A_{\mu}$.

Corresponding scattering cross-sections for spin-dependent (SD) and spin-independent (SI) elastic scattering $\chi u \rightarrow \chi u$ are given by formulas (3.6 --- 3.11) of \cite{difranzo2013simplified}.

\begin{figure}[h]
   \begin{minipage}{0.33\textwidth}
     \centering
     \includegraphics[width=.99\linewidth]{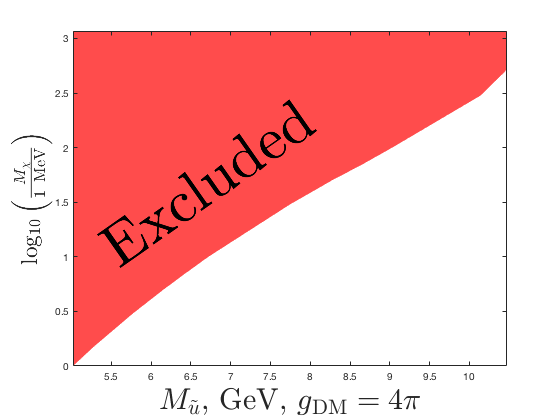}
   \end{minipage}\hfill
   \begin{minipage}{0.33\textwidth}
     \centering
     \includegraphics[width=.99\linewidth]{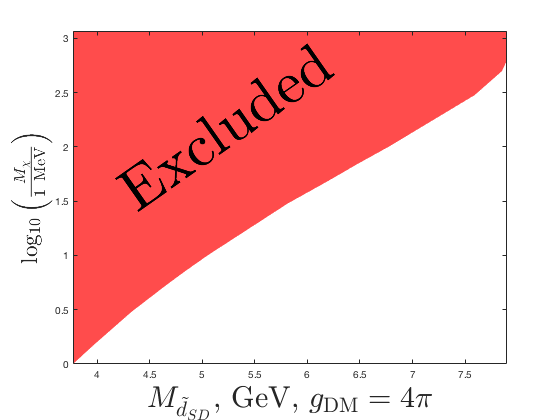}
   \end{minipage}\hfill
   \begin{minipage}{0.33\textwidth}
     \centering
     \includegraphics[width=.99\linewidth]{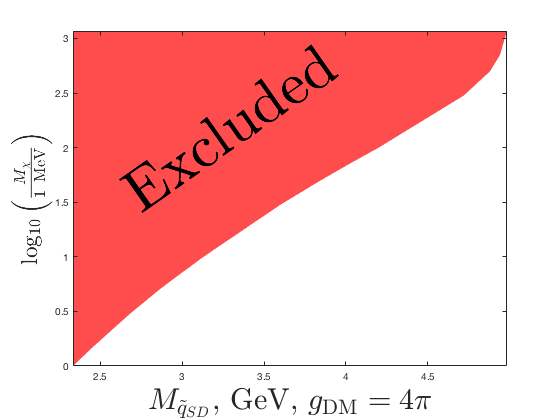}
   \end{minipage}
   \begin{minipage}{0.33\textwidth}
     \centering
     \includegraphics[width=.99\linewidth]{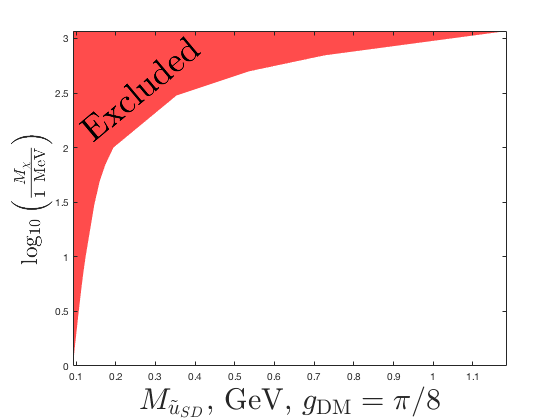}
   \end{minipage}\hfill
   \begin{minipage}{0.33\textwidth}
     \centering
     \includegraphics[width=.99\linewidth]{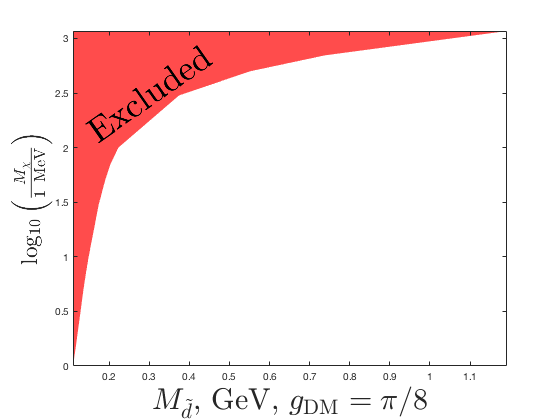}
   \end{minipage}\hfill
   \begin{minipage}{0.33\textwidth}
     \centering
     \includegraphics[width=.99\linewidth]{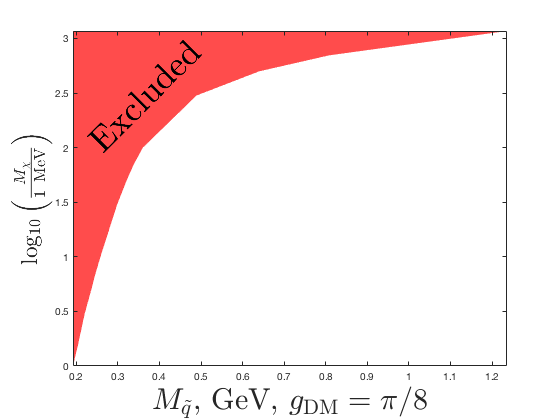}
   \end{minipage}
   \caption{Constraints on mediator particle mass $(M_{\tilde{u}}, M_{\tilde{d}}, M_{\tilde{q}})$ and DM SIMP mass $M_{\chi}$}\label{SIMP_constraints_1}.
\end{figure}

Substituting the scattering cross sections into Eq.\ref{simps_equation}, we obtained constraints on the parameters of this model $(M_{\chi}, M_{\tilde{u}}, M_{\tilde{q}}, M_{\tilde{d}})$ for various interaction constants $g_{DM}$ shown in Fig.~\ref{SIMP_constraints_1}.

We considered also the model described in the work of N. Daci, I. De Bruyn et al.\ \cite{daci2015simplified} and the similar model used by Patrick J. Fox, Roni Harnik et al.\ in \cite{fox2012missing}, where SIMP-baryon scattering cross-section has a form:

\begin{equation}
    \sigma_{\chi N} \approx {g_{\chi}^2 g_q^2 \over \pi m_{\phi}^4} \mu_{\chi N}^2 f_N^2.
\end{equation}

\begin{figure}[h]
  \begin{minipage}{0.33\textwidth}
     \centering
     \includegraphics[width=.99\linewidth]{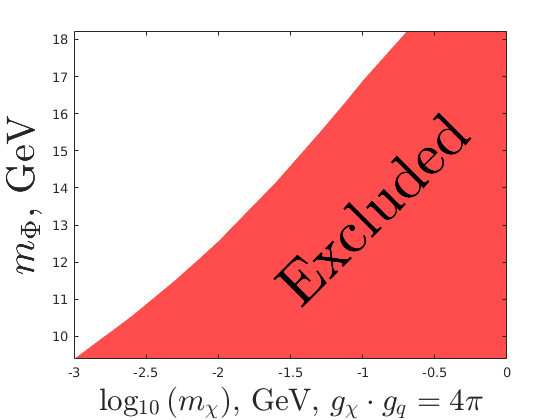}
  \end{minipage}\hfill
  \begin{minipage}{0.33\textwidth}
     \centering
     \includegraphics[width=.99\linewidth]{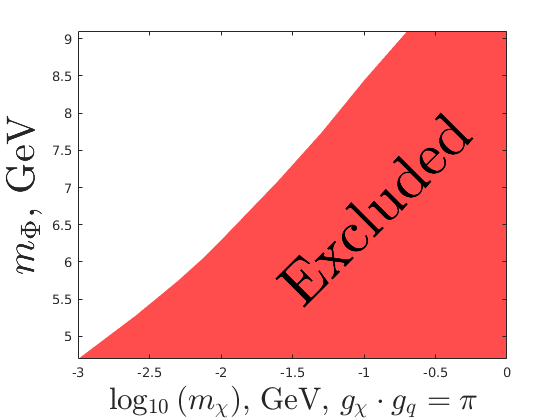}
    \end{minipage}\hfill
  \begin{minipage}{0.33\textwidth}
     \centering
     \includegraphics[width=.99\linewidth]{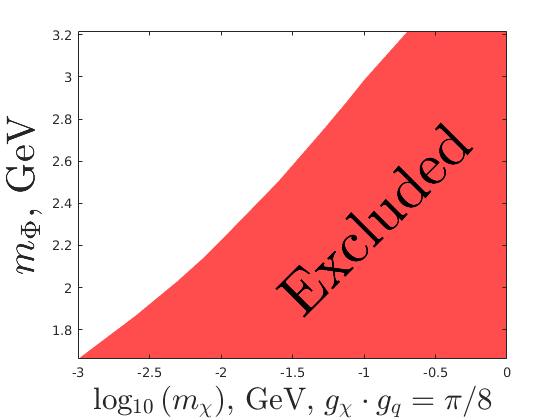}
  \end{minipage}
  \caption{Constraints on mediator's particle mass $(m_{\Phi})$ and SIMP mass $m_{\chi}$}
  \label{simplified_simps_and_lhc}
\end{figure}

Substituting again this scattering cross section into Eq.~\ref{simps_equation} one can obtain the constraints on masses of SIMP $m_{\chi}$ and mediator masses $m_{\Phi}$ (see Fig.~\ref{simplified_simps_and_lhc}).

The similar results are obtained for DM mass and Energy scale $\Lambda$ (see Eq.~\ref{Kenji_cross_section}) for Kenji Kadota's model \cite{kadota2012effects} where a simple model was studied in which spin independent scattering cross-section is:
\begin{equation}
    \sigma_{SI} \approx {{4 \mu_p^2} \over {\pi \Lambda^6}} f_p^2, 
    \label{Kenji_cross_section}
\end{equation}
with the nucleon-DM reduced mass $\mu$ and effective coupling of the DM with the nucleon $f_p$. 

Our approach to constraining the parameters of particular SIMP models, which we are considered here, may be generalized for any other similar models.

\section{Other constraints}
\label{Other}
Here we list other possible indirect DM signals, non-observation of which puts constraint on DM properties.

There are plenty of works about neutrino-based constraints on DM. The majority of such constraints comes from neutrino fluxes from the Sun and the core of the Earth. The DM particles can accumulate there, producing the neutrinos via decay or annihilation and therefore changing their expected flux. Accumulation is supposed to proceed due to interaction of particles of DM and ordinary matter.

Many neutrino experiments, such as IceCube, Super-Kamiokanda etc, are aimed to catch this deviation. However, it still is not found, and these experiments just set the constraints on the DM properties \cite{2018arXiv180403848I, 2019arXiv190712509I, 2019arXiv190711699I}.

Effects of accumulation can take place even if DM particles both interact with ordinary ones and do not. It may happen in case of self-interacting DM \cite{2019JCAP...06..022G}. Given so, the particles in gravitational field of the Sun or the Earth can collide with each other what will change their capture rate. It can be, that for the Earth it will diminish rate, while for the Sun it will increase.

Another specific model case which may give indirect observational DM effect can appear when DM consists of WIMP and Primordial Black Holes (PBH) under assumption of existence of the clumps. In this case PBH should increase annihilation signal so that some parameters of such model can be already excluded by observation of gamma-radiation \cite{Eroshenko2016, 2019PhRvD.100b3506A}.

\section{Conclusion}
\label{Conclusion}

In this article we briefly review a set of possible indirect effects of DM. It does not intend to be complete, more attention is paid to the effects which were a subject of our investigations.

First of all we consider effects in cosmic rays. We demonstrate that observed anomalies in high-energy cosmic positrons and electrons cannot be explained by annihilation or decay of dark matter in the framework of usual assumption on DM distribution in space and their kinematics and physics without conflicting with data on gamma-ray background (IGRB) and CMB. We show that CMB constraint can be circumvented under additional model assumption, while IGRB cannot and it provides a least model-dependent constraint of DM interpretation of CR data. The possible exceptions are Dark disk model and maybe clumps. However, we show that DM clumps are within sensitivity of existing gamma-telescopes. Many kinematical and physical variations of DM models are considered (cascade processes, different Lagrangians) and do not not give us positive results. The only hope which we see at given moment is the effect of the identical fermions in the final state.

Another part of this mini-review is devoted to SIMPs. Effect of Silk-like damping of large inhomogeneities of matter in the Universe allows constraining SIMP-baryon cross-section depending on SIMP mass in a simple manner. This effect has been considered relatively recently in application to DM and it gives stronger restriction on DM properties as compared to previously more commonly used one, based on damping effect due to free-streaming, diffusion and other. It allows constraining the model. We do it in generic case and apply it to several specific models.

Also, we list some other possible indirect DM effects, some of which require extra model assumptions.

To sum up, at present time many expected manifestations of DM are not observed, while observed phenomena of unknown nature cannot be definitely connected with dark matter. But alternative hypotheses (e.g. pulsar explanation of electron-positron excesses in cosmic rays) suffer, at least, the same. So, we do not claim that possibility of DM indirect effects observation is excluded.

\section{Acknowledgment}
We would like to thank R.I.Budaev, M.N.Laletin for background created by them to this work, V.A.Lensky for theoretical support, M.Yu.Khlopov, S.G.Rubin for interest to the work with useful discussion and A.A.Kirillov for technical assistance. 
Special thanks we would like to address to V.A.Rubakov, A.Loeb and E.Berchinger for the help in clarifying the case with Silk-like effect of dark matter.

The work 
was supported by the Ministry of Education and Science of the Russian Federation, MEPhI Academic Excellence Project (contract № 02.a03.21.0005, 27.08.2013). The work of K.B. is also funded by the Ministry of Education and Science of the Russia, Project № 3.6760.2017/BY. The work of E.E. was supported by fund Basis, project № 18-1-5-89-1.

\appendix
\section*{Appendix}
\label{Appendix}

Here we obtain matrix element for reactions with final identiacal fermions. 
In terms of quantum field theory, the case when the initial state is fixed  can be expressed as $\wick{ \c1 {X} |\c1 {p}>}$.

To construct the corresponding matrix elements, we first use the second quantization of Dirac spinors \cite{peskin2018introduction}

\begin{equation}
    \Psi(x) = \int {d^3p \over (2\pi)^3}{1 \over \sqrt{2 E_p}} \sum_s \left(a_p^s  u^s(p) e^{-ipx} + {b_p^s}^\dagger v^s(p) e^{ipx}\right),
\end{equation}
\begin{equation}
    \overline{\Psi}(y) = \int {d^3q \over (2\pi)^3}{1 \over \sqrt{2 E_q}} \sum_r \left(b_q^r \bar{v^r}(q) e^{-iqy} + {a_q^r}^\dagger \bar{u^r}(q) e^{iqy}\right),
\end{equation}
where $\overline{\Psi^C} = (-i\gamma^2 \Psi^*)^\dagger \gamma^0 = \Psi^T (-i \gamma^2)^T \gamma^0 = \Psi^T C^T \gamma^0$.
For example, the matrix element of two-particle decay (Fig. \ref{feyn3}) in this case is given as:
\begin{equation}
\begin{gathered}
M = \wick{
 < \c2 {k_1} \c1 {k_2} | \c3 {X}  \overline{\c2 {\Psi}}^C (a+b \gamma^5) \c1 {\Psi} | \c3 {p}>
} -
\wick{
 < \c1 {k_1} \c2 {k_2} | \c3 {X}  \overline{\c2 {\Psi}}^C (a+b \gamma^5) \c1 {\Psi} | \c3 {p}>
} = \\
= \wick{
 < \c2 {k_1} \c1 {k_2} | \c3 {X}  \c2 {\Psi}^T C^T \gamma^0 (a+b \gamma^5) \c1 {\Psi} | \c3 {p}>
} -
\wick{
 < \c1 {k_1} \c2 {k_2} | \c3 {X}  \c2 {\Psi}^T C^T \gamma^0 (a+b \gamma^5) \c1 {\Psi} | \c3 {p}>
} = \\
= v^T(k_1) C^T \gamma^0 (a+b \gamma^5) v(k_2) - v^T(k_2) C^T \gamma^0 (a+b \gamma^5) v(k_1) .
\end{gathered}
\end{equation}

\bibliographystyle{JHEP}
\bibliography{Bibliography}

\providecommand{\href}[2]{#2}\begingroup\raggedright\begin{thebibliography}{10}

\bibitem{Zeldovich}
Y.~B.~Zeldovich, A.~Klypin, M.~Khlopov and V.~M.~Chechetkin,
  \emph{Astrophysical bounds on the mass of heavy stable neutral leptons},
  {\emph{Sov. J. Nucl. Phys} {\bf 31} (05, 1980) 664--669}.

\bibitem{Konoplich}
R.~V.~Konoplich and M.~Yu.~Khlopov, \emph{Astrophysical constraints on mass of
  very heavy stable neutrino}, {\emph{Phys. Atom. Nucl.} {\bf 57} (1994)
  425--431}.

\bibitem{PhysRevLett.63.840}
A.~J. Tylka, \emph{Cosmic-ray positrons from annihilation of weakly interacting
  massive particles in the galaxy},
  \href{http://dx.doi.org/10.1103/PhysRevLett.63.840}{\emph{Phys. Rev. Lett.}
  {\bf 63} (Aug, 1989) 840--843}.

\bibitem{PhysRevD.42.1001}
M.~S. Turner and F.~Wilczek, \emph{Positron line radiation as a signature of
  particle dark matter in the halo},
  \href{http://dx.doi.org/10.1103/PhysRevD.42.1001}{\emph{Phys. Rev. D} {\bf
  42} (Aug, 1990) 1001--1007}.

\bibitem{PhysRevD.43.1774}
M.~Kamionkowski and M.~S. Turner, \emph{Distinctive positron feature from
  particle dark-matter annihilations in the galatic halo},
  \href{http://dx.doi.org/10.1103/PhysRevD.43.1774}{\emph{Phys. Rev. D} {\bf
  43} (Mar, 1991) 1774--1780}.

\bibitem{Adriani:2008zr}
{\scshape PAMELA} collaboration, O.~Adriani et~al., \emph{{An anomalous
  positron abundance in cosmic rays with energies 1.5-100 GeV}},
  \href{http://dx.doi.org/10.1038/nature07942}{\emph{Nature} {\bf 458} (2009)
  607--609}, [\href{http://arxiv.org/abs/0810.4995}{{\tt 0810.4995}}].

\bibitem{PhysRevLett.110.141102}
{\scshape AMS Collaboration} collaboration, M.~Aguilar, G.~Alberti, B.~Alpat,
  A.~Alvino, G.~Ambrosi, K.~Andeen et~al., \emph{First result from the alpha
  magnetic spectrometer on the international space station: Precision
  measurement of the positron fraction in primary cosmic rays of 0.5-350 gev},
  \href{http://dx.doi.org/10.1103/PhysRevLett.110.141102}{\emph{Phys. Rev.
  Lett.} {\bf 110} (Apr, 2013) 141102}.

\bibitem{PhysRevLett.113.121101}
{\scshape AMS Collaboration} collaboration, L.~Accardo, M.~Aguilar, D.~Aisa,
  B.~Alpat, A.~Alvino, G.~Ambrosi et~al., \emph{High statistics measurement of
  the positron fraction in primary cosmic rays of 0.5--500 gev with the alpha
  magnetic spectrometer on the international space station},
  \href{http://dx.doi.org/10.1103/PhysRevLett.113.121101}{\emph{Phys. Rev.
  Lett.} {\bf 113} (Sep, 2014) 121101}.

\bibitem{2009PhRvL.103e1101Y}
H.~{Y{\"u}ksel}, M.~D. {Kistler} and T.~{Stanev}, \emph{{TeV Gamma Rays from
  Geminga and the Origin of the GeV Positron Excess}},
  \href{http://dx.doi.org/10.1103/PhysRevLett.103.051101}{\emph{Physical Review
  Letters} {\bf 103} (Jul, 2009) 051101},
  [\href{http://arxiv.org/abs/0810.2784}{{\tt 0810.2784}}].

\bibitem{PhysRevD.52.3265}
A.~M. Atoyan, F.~A. Aharonian and H.~J. V\"olk, \emph{Electrons and positrons
  in the galactic cosmic rays},
  \href{http://dx.doi.org/10.1103/PhysRevD.52.3265}{\emph{Phys. Rev. D} {\bf
  52} (Sep, 1995) 3265--3275}.

\bibitem{2019PhRvD..99j3022R}
S.~{Recchia}, S.~{Gabici}, F.~A. {Aharonian} and J.~{Vink}, \emph{{Local fading
  accelerator and the origin of TeV cosmic ray electrons}},
  \href{http://dx.doi.org/10.1103/PhysRevD.99.103022}{\emph{Phys. Rev. D} {\bf
  99} (May, 2019) 103022}, [\href{http://arxiv.org/abs/1811.07551}{{\tt
  1811.07551}}].

\bibitem{2018PhRvD..97f3011K}
M.~{Kachelrie{\ss}}, A.~{Neronov} and D.~V. {Semikoz}, \emph{{Cosmic ray
  signatures of a 2-3 Myr old local supernova}},
  \href{http://dx.doi.org/10.1103/PhysRevD.97.063011}{\emph{Phys. Rev. D} {\bf
  97} (Mar, 2018) 063011}, [\href{http://arxiv.org/abs/1710.02321}{{\tt
  1710.02321}}].

\bibitem{2019arXiv190206173L}
P.~{Lipari}, \emph{{Understanding the cosmic ray positron flux}}, {\emph{arXiv
  e-prints} (Feb, 2019) arXiv:1902.06173},
  [\href{http://arxiv.org/abs/1902.06173}{{\tt 1902.06173}}].

\bibitem{2015ICRC...34...14C}
M.~{Cirelli}, \emph{{Dark matter phenomena}},  in \emph{34th International
  Cosmic Ray Conference (ICRC2015)}, vol.~34, p.~14, Jul, 2015.
\newblock \href{http://arxiv.org/abs/1511.02031}{{\tt 1511.02031}}.

\bibitem{Ackermann:2014usa}
{\scshape Fermi-LAT} collaboration, M.~Ackermann et~al., \emph{{The spectrum of
  isotropic diffuse gamma-ray emission between 100 MeV and 820 GeV}},
  \href{http://dx.doi.org/10.1088/0004-637X/799/1/86}{\emph{Astrophys. J.} {\bf
  799} (2015) 86}, [\href{http://arxiv.org/abs/1410.3696}{{\tt 1410.3696}}].

\bibitem{Belotsky:2016tja}
K.~Belotsky, R.~Budaev, A.~Kirillov and M.~Laletin, \emph{{Fermi-LAT kills dark
  matter interpretations of AMS-02 data. Or not?}},
  \href{http://dx.doi.org/10.1088/1475-7516/2017/01/021}{\emph{JCAP} {\bf 1701}
  (2017) 021}, [\href{http://arxiv.org/abs/1606.01271}{{\tt 1606.01271}}].

\bibitem{Fowlie:2017fya}
A.~Fowlie, \emph{{DAMPE squib? Significance of the 1.4 TeV DAMPE excess}},
  \href{http://dx.doi.org/10.1016/j.physletb.2018.03.006}{\emph{Phys. Lett.}
  {\bf B780} (2018) 181--184}, [\href{http://arxiv.org/abs/1712.05089}{{\tt
  1712.05089}}].

\bibitem{Cao:2017rjr}
J.~Cao, X.~Guo, L.~Shang, F.~Wang, P.~Wu and L.~Zu, \emph{{Scalar dark matter
  explanation of the DAMPE data in the minimal Left-Right symmetric model}},
  \href{http://dx.doi.org/10.1103/PhysRevD.97.063016}{\emph{Phys. Rev.} {\bf
  D97} (2018) 063016}, [\href{http://arxiv.org/abs/1712.05351}{{\tt
  1712.05351}}].

\bibitem{Liu:2017obm}
G.-L. Liu, F.~Wang, W.~Wang and J.~M. Yang, \emph{{Explaining DAMPE results by
  dark matter with hierarchical lepton-specific Yukawa interactions}},
  \href{http://dx.doi.org/10.1088/1674-1137/42/3/035101}{\emph{Chin. Phys.}
  {\bf C42} (2018) 035101}, [\href{http://arxiv.org/abs/1712.02381}{{\tt
  1712.02381}}].

\bibitem{Ding:2017jdr}
R.~Ding, Z.-L. Han, L.~Feng and B.~Zhu, \emph{{Confronting the DAMPE Excess
  with the Scotogenic Type-II Seesaw Model}},
  \href{http://dx.doi.org/10.1088/1674-1137/42/8/083104}{\emph{Chin. Phys.}
  {\bf C42} (2018) 083104}, [\href{http://arxiv.org/abs/1712.02021}{{\tt
  1712.02021}}].

\bibitem{Li:2017tmd}
T.~Li, N.~Okada and Q.~Shafi, \emph{{Scalar dark matter, Type II Seesaw and the
  DAMPE cosmic ray $e^+ + e^-$ excess}},
  \href{http://dx.doi.org/10.1016/j.physletb.2018.02.006}{\emph{Phys. Lett.}
  {\bf B779} (2018) 130--135}, [\href{http://arxiv.org/abs/1712.00869}{{\tt
  1712.00869}}].

\bibitem{Chen:2017tva}
C.-H. Chen, C.-W. Chiang and T.~Nomura, \emph{{Explaining the DAMPE $e^+ e^-$
  excess using the Higgs triplet model with a vector dark matter}},
  \href{http://dx.doi.org/10.1103/PhysRevD.97.061302}{\emph{Phys. Rev.} {\bf
  D97} (2018) 061302}, [\href{http://arxiv.org/abs/1712.00793}{{\tt
  1712.00793}}].

\bibitem{2019PDU....2600333B}
K.~{Belotsky}, A.~{Kamaletdinov}, M.~{Laletin} and M.~{Solovyov}, \emph{{The
  DAMPE excess and gamma-ray constraints}},
  \href{http://dx.doi.org/10.1016/j.dark.2019.100333}{\emph{Physics of the Dark
  Universe} {\bf 26} (Dec, 2019) 100333},
  [\href{http://arxiv.org/abs/1904.02456}{{\tt 1904.02456}}].

\bibitem{ade2016planck}
P.~A. Ade, N.~Aghanim, M.~Arnaud, M.~Ashdown, J.~Aumont, C.~Baccigalupi et~al.,
  \emph{Planck 2015 results-xiii. cosmological parameters}, {\emph{Astronomy \&
  Astrophysics} {\bf 594} (2016) A13}.

\bibitem{xiang2017dark}
Q.-F. Xiang, X.-J. Bi, S.-J. Lin and P.-F. Yin, \emph{A dark matter model that
  reconciles tensions between the cosmic-ray e$\pm$excess and the gamma-ray and
  cmb constraints}, {\emph{Physics Letters B} {\bf 773} (2017) 448--454}.

\bibitem{ibe2009breit}
M.~Ibe, H.~Murayama and T.~Yanagida, \emph{Breit-wigner enhancement of dark
  matter annihilation}, {\emph{Physical Review D} {\bf 79} (2009) 095009}.

\bibitem{bai2018supersymmetric}
Y.~Bai, J.~Berger and S.~Lu, \emph{Supersymmetric resonant dark matter: a
  thermal model for the ams-02 positron excess}, {\emph{Physical Review D} {\bf
  97} (2018) 115012}.

\bibitem{diamanti2014constraining}
R.~Diamanti, L.~Lopez-Honorez, O.~Mena, S.~Palomares-Ruiz and A.~C. Vincent,
  \emph{Constraining dark matter late-time energy injection: decays and p-wave
  annihilations}, {\emph{Journal of Cosmology and Astroparticle Physics} {\bf
  2014} (2014) 017}.

\bibitem{buch2017late}
J.~Buch, P.~Ralegankar and V.~Rentala, \emph{Late decaying 2-component dark
  matter scenario as an explanation of the ams-02 positron excess},
  {\emph{Journal of Cosmology and Astroparticle Physics} {\bf 2017} (2017)
  028}.

\bibitem{KB2000}
K.~Belotsky, M.~Khlopov and K.~Shibaev, \emph{Sakharov's enhancement in the
  effect of 4th generation neutrino}, {\emph{Grav. Cosmol. Suppl.} {\bf 6} (01,
  2000) 140--143}.

\bibitem{2008PAN....71..147B}
K.~M. {Belotsky}, D.~{Fargion}, M.~Y. {Khlopov} and R.~V. {Konoplich},
  \emph{{May heavy neutrinos solve underground and cosmic-ray puzzles?}},
  \href{http://dx.doi.org/10.1007/s11450-008-1016-9}{\emph{Physics of Atomic
  Nuclei} {\bf 71} (Jan, 2008) 147--161},
  [\href{http://arxiv.org/abs/hep-ph/0411093}{{\tt hep-ph/0411093}}].

\bibitem{2005GrCo...11...27B}
K.~M. {Belotsky}, M.~Y. {Khlopov}, S.~V. {Legonkov} and K.~I. {Shibaev},
  \emph{{Effects of a new long-range interaction: Recombination of relic heavy
  neutrinos and antineutrinos}}, {\emph{Gravitation and Cosmology} {\bf 11}
  (Jun, 2005) 27--33}, [\href{http://arxiv.org/abs/astro-ph/0504621}{{\tt
  astro-ph/0504621}}].

\bibitem{2009PhRvD..79a5014A}
N.~{Arkani-Hamed}, D.~P. {Finkbeiner}, T.~R. {Slatyer} and N.~{Weiner},
  \emph{{A theory of dark matter}},
  \href{http://dx.doi.org/10.1103/PhysRevD.79.015014}{\emph{Phys. Rev. D} {\bf
  79} (Jan, 2009) 015014}, [\href{http://arxiv.org/abs/0810.0713}{{\tt
  0810.0713}}].

\bibitem{Belotsky:2017wgi}
K.~M. Belotsky, R.~I. Budaev, A.~A. Kirillov and M.~L. Solovyov,
  \emph{{Gamma-rays from possible disk component of dark matter}},
  \href{http://dx.doi.org/10.1088/1742-6596/798/1/012084}{\emph{J. Phys. Conf.
  Ser.} {\bf 798} (2017) 012084}.

\bibitem{1742-6596-675-1-012023}
V.~V. Alekseev, K.~M. Belotsky, Y.~V. Bogomolov, R.~I. Budaev, O.~A. Dunaeva,
  A.~A. Kirillov et~al., \emph{High-energy cosmic antiparticle excess vs.
  isotropic gamma-ray background problem in decaying dark matter universe},
  {\emph{Journal of Physics: Conference Series} {\bf 675} (2016) 012023}.

\bibitem{1742-6596-675-1-012026}
V.~V. Alekseev, K.~M. Belotsky, Y.~V. Bogomolov, R.~I. Budaev, O.~A. Dunaeva,
  A.~A. Kirillov et~al., \emph{On a possible solution to gamma-ray
  overabundance arising in dark matter explanation of cosmic antiparticle
  excess}, {\emph{Journal of Physics: Conference Series} {\bf 675} (2016)
  012026}.

\bibitem{Alekseev2017An}
V.~V. Alekseev, K.~M. Belotsky, Y.~V. Bogomolov, R.~I. Budaev, O.~A. Dunaeva,
  A.~A. Kirillov et~al., \emph{Analysis of a possible explanation of the
  positron anomaly in terms of dark matter},
  \href{http://dx.doi.org/10.1134/S1063778817040020}{\emph{Physics of Atomic
  Nuclei} {\bf 80} (Jul, 2017) 713--717}.

\bibitem{Belotsky:2018vyt}
K.~M. Belotsky, A.~A. Kirillov and M.~L. Solovyov, \emph{{Development of dark
  disk model of positron anomaly origin}},
  \href{http://dx.doi.org/10.1142/S0218271818410109}{\emph{Int. J. Mod. Phys.}
  {\bf D27} (2018) 1841010}, [\href{http://arxiv.org/abs/1802.04678}{{\tt
  1802.04678}}].

\bibitem{FermiLAT:performance}
``{Fermi-LAT performance}.''
  \url{http://www.slac.stanford.edu/exp/glast/groups/canda/lat_Performance.htm}.

\bibitem{PointSourceSensitivities}
P.~Di~Sciascio, ``{Roundtable on New High-Altitude Experiment in the South,
  Synergy with CTA}.''
  \url{http://www.ssdc.asi.it/16thagilemeeting/dwl.php?file=workshop_files/slide/17a-DiSciascio_16th_Agile_WS.pdf},
  2018.

\bibitem{belyaev2013calchep}
A.~Belyaev, N.~D. Christensen and A.~Pukhov, \emph{Calchep 3.4 for collider
  physics within and beyond the standard model}, {\emph{Computer Physics
  Communications} {\bf 184} (2013) 1729--1769}.

\bibitem{Khlopov2006}
M.~Y. Khlopov, \emph{Composite dark matter from the fourth generation},
  \href{http://dx.doi.org/10.1134/S0021364006010012}{\emph{JETP Letters} {\bf
  83} (2006) 1--4}.

\bibitem{khlopov2006dark}
M.~Y. Khlopov, C.~Stephan and D.~Fargion, \emph{Dark matter with invisible
  light from heavy double charged leptons of almost-commutative geometry?},
  {\emph{Classical and Quantum Gravity} {\bf 23} (2006) 7305}.

\bibitem{khlopov2008strong}
M.~Y. Khlopov and C.~Kouvaris, \emph{Strong interactive massive particles from
  a strong coupled theory}, {\emph{Physical Review D} {\bf 77} (2008) 065002}.

\bibitem{doi:10.1142/S0217751X14430027}
M.~Khlopov, \emph{Dark atoms and puzzles of dark matter searches},
  \href{http://dx.doi.org/10.1142/S0217751X14430027}{\emph{International
  Journal of Modern Physics A} {\bf 29} (2014) 1443002},
  [\href{http://arxiv.org/abs/http://www.worldscientific.com/doi/pdf/10.1142/S0217751X14430027}{{\tt
  http://www.worldscientific.com/doi/pdf/10.1142/S0217751X14430027}}].

\bibitem{Belotsky:2014haa}
K.~Belotsky, M.~Khlopov, C.~Kouvaris and M.~Laletin, \emph{{Decaying Dark Atom
  constituents and cosmic positron excess}},
  \href{http://dx.doi.org/10.1155/2014/214258}{\emph{Adv. High Energy Phys.}
  {\bf 2014} (2014) 214258}, [\href{http://arxiv.org/abs/1403.1212}{{\tt
  1403.1212}}].

\bibitem{belotsky2014dark}
K.~Belotsky, M.~Khlopov and M.~Laletin, \emph{Dark atoms and their decaying
  constituents}, {\emph{arXiv preprint arXiv:1411.3657} (2014) }.

\bibitem{brown1995understanding}
R.~W. Brown, \emph{Understanding something about nothing: radiation zeros},  in
  \emph{AIP Conference Proceedings}, vol.~350, pp.~261--272, AIP, 1995.

\bibitem{bertschinger2006effects}
E.~Bertschinger, \emph{Effects of cold dark matter decoupling and pair
  annihilation on cosmological perturbations}, {\emph{Physical Review D} {\bf
  74} (2006) 063509}.

\bibitem{loeb2005small}
A.~Loeb and M.~Zaldarriaga, \emph{Small-scale power spectrum of cold dark
  matter}, {\emph{Physical Review D} {\bf 71} (2005) 103520}.

\bibitem{ooba2019cosmological}
J.~Ooba, H.~Tashiro and K.~Kadota, \emph{Cosmological constraints on the
  velocity-dependent baryon-dark matter coupling}, {\emph{arXiv preprint
  arXiv:1902.00826} (2019) }.

\bibitem{belotsky2016temperature}
K.~Belotsky, E.~Esipova and A.~Kirillov, \emph{On the temperature evolution of
  multicomponent dark matter with coulomb-like interaction},  in \emph{Journal
  of Physics: Conference Series}, vol.~675, p.~012017, IOP Publishing, 2016.

\bibitem{difranzo2013simplified}
A.~DiFranzo, K.~I. Nagao, A.~Rajaraman and T.~M. Tait, \emph{Simplified models
  for dark matter interacting with quarks}, {\emph{Journal of High Energy
  Physics} {\bf 2013} (2013) 14}.

\bibitem{daci2015simplified}
N.~Daci, I.~De~Bruyn, S.~Lowette, M.~H. Tytgat and B.~Zaldivar,
  \emph{Simplified simps and the lhc}, {\emph{Journal of High Energy Physics}
  {\bf 2015} (2015) 108}.

\bibitem{fox2012missing}
P.~J. Fox, R.~Harnik, J.~Kopp and Y.~Tsai, \emph{Missing energy signatures of
  dark matter at the lhc}, {\emph{Physical Review D} {\bf 85} (2012) 056011}.

\bibitem{kadota2012effects}
K.~Kadota, \emph{The effects of quark interactions on the dark matter kinetic
  decoupling}, {\emph{PoS} (2012) 029}.

\bibitem{2018arXiv180403848I}
{IceCube Collaboration}, M.~G. {Aartsen}, M.~{Ackermann}, J.~{Adams}, J.~A.
  {Aguilar}, M.~{Ahlers} et~al., \emph{{Search for neutrinos from decaying dark
  matter with IceCube}}, {\emph{arXiv e-prints} (Apr, 2018) arXiv:1804.03848},
  [\href{http://arxiv.org/abs/1804.03848}{{\tt 1804.03848}}].

\bibitem{2019arXiv190712509I}
{IceCube Collaboration}, M.~G. {Aartsen}, M.~{Ackermann}, J.~{Adams}, J.~A.
  {Aguilar}, M.~{Ahlers} et~al., \emph{{Velocity independent constraints on
  spin-dependent DM-nucleon interactions from IceCube and PICO}}, {\emph{arXiv
  e-prints} (Jul, 2019) arXiv:1907.12509},
  [\href{http://arxiv.org/abs/1907.12509}{{\tt 1907.12509}}].

\bibitem{2019arXiv190711699I}
{IceCube Collaboration}, M.~G. {Aartsen}, M.~{Ackermann}, J.~{Adams}, J.~A.
  {Aguilar}, M.~{Ahlers} et~al., \emph{{The IceCube Neutrino Observatory --
  Contributions to the 36th International Cosmic Ray Conference (ICRC2019)}},
  {\emph{arXiv e-prints} (Jul, 2019) arXiv:1907.11699},
  [\href{http://arxiv.org/abs/1907.11699}{{\tt 1907.11699}}].

\bibitem{2019JCAP...06..022G}
C.~{Gaidau} and J.~{Shelton}, \emph{{A solar system test of self-interacting
  dark matter}},
  \href{http://dx.doi.org/10.1088/1475-7516/2019/06/022}{\emph{JCAP} {\bf 2019}
  (Jun, 2019) 022}, [\href{http://arxiv.org/abs/1811.00557}{{\tt 1811.00557}}].

\bibitem{Eroshenko2016}
Y.~N. Eroshenko, \emph{Dark matter density spikes around primordial black
  holes}, \href{http://dx.doi.org/10.1134/S1063773716060013}{\emph{Astronomy
  Letters} {\bf 42} (Jun, 2016) 347--356}.

\bibitem{2019PhRvD.100b3506A}
J.~{Adamek}, C.~T. {Byrnes}, M.~{Gosenca} and S.~{Hotchkiss}, \emph{{WIMPs and
  stellar-mass primordial black holes are incompatible}},
  \href{http://dx.doi.org/10.1103/PhysRevD.100.023506}{\emph{Physical Rewiev D}
  {\bf 100} (Jul, 2019) 023506}, [\href{http://arxiv.org/abs/1901.08528}{{\tt
  1901.08528}}].

\bibitem{peskin2018introduction}
M.~E. Peskin, \emph{An introduction to quantum field theory}.
\newblock CRC Press, 2018.

\end{thebibliography}\endgroup

\end{document}